\documentclass[twocolumn,prb,superscriptaddress]{revtex4-1}

\usepackage{graphicx}
\usepackage{dcolumn}
\usepackage{hyperref}
\usepackage{amsmath}
\usepackage{amssymb}
\usepackage{color}
\usepackage{txfonts}
\begin{document}
\title{Optical addressing of an individual erbium ion in silicon}
\small
\author{Chunming Yin}
\affiliation{Centre of Excellence for Quantum Computation and Communication Technology, School of Physics, University of New South Wales, Sydney, NSW 2052, Australia.}
\author{Milos Rancic}
\affiliation{Centre of Excellence for Quantum Computation and Communication Technology, RSPE, Australian National University, Canberra, ACT 0200, Australia}
\author{Gabriele G. de Boo}
\affiliation{Centre of Excellence for Quantum Computation and Communication Technology, School of Physics, University of New South Wales, Sydney, NSW 2052, Australia.}
\author{Nikolas Stavrias}
\affiliation{Centre of Excellence for Quantum Computation and Communication Technology, School of Physics, University of Melbourne, Melbourne, VIC 3010, Australia}
\author{Jeffrey C. McCallum}
\affiliation{Centre of Excellence for Quantum Computation and Communication Technology, School of Physics, University of Melbourne, Melbourne, VIC 3010, Australia}
\author{Matthew J. Sellars}
\affiliation{Centre of Excellence for Quantum Computation and Communication Technology, RSPE, Australian National University, Canberra, ACT 0200, Australia}
\author{Sven Rogge}
\affiliation{Centre of Excellence for Quantum Computation and Communication Technology, School of Physics, University of New South Wales, Sydney, NSW 2052, Australia.}
             
\maketitle
The detection of electron spins associated with single defects in solids is a critical operation for a range of quantum information and measurement applications currently under development\cite{morello_single-shot_2010,fuechsle_single-atom_2012,pla_single-atom_2012,gaebel_room-temperature_2006,jiang_repetitive_2009,togan_quantum_2010,maze_nanoscale_2008,morton_embracing_2011,zwanenburg_silicon_2012}. To date, it has only been accomplished for two centres in crystalline solids: phosphorus in silicon using electrical readout based on a single electron transistor (SET)\cite{morello_single-shot_2010} and nitrogen-vacancy centres in diamond using optical readout\cite{gaebel_room-temperature_2006,togan_quantum_2010}. A spin readout fidelity of about 90\% has been demonstrated with both electrical readout\cite{morello_single-shot_2010} and optical readout\cite{robledo_high-fidelity_2011,neumann_single-shot_2010}, however, the thermal limitations of the electrical readout and the poor photon collection efficiency of the optical readout hinder achieving the high fidelity required for quantum information applications. Here we demonstrate a hybrid approach using optical excitation to change the charge state of the defect centre in a silicon-based SET, conditional on its spin state, and then detecting this change electrically. The optical frequency addressing in high spectral resolution conquers the thermal broadening limitation of the previous electrical readout and charge sensing avoids the difficulties of efficient photon collection. This is done with erbium in silicon and has the potential to enable new architectures for quantum information processing devices and to dramatically increase the range of defect centres that can be exploited. Further, the efficient electrical detection of the optical excitation of single sites in silicon is a major step in developing an interconnect between silicon and optical based quantum computing technologies.

The potential for a hybrid optical/electrical single-spin readout was recently established by Steger et al., revealing long nuclear spin coherence time with an ensemble of P ions in highly purified $^{28}$Si\cite{steger_quantum_2012}. The readout of the spin ensemble was demonstrated by detecting the photocurrent generated when the excitonic transition at 1,078 nm associated with the P ions was excited. In the current work we demonstrate single-site detection by electrically detecting the optical excitation of the $^{4}$I$_{15/2}$ - $^{4}$I$_{13/2}$ transition of single Er ions implanted into silicon, resolving both electronic Zeeman and hyperfine structure. The efficient readout required for single-site detection is achieved by measuring the photo-induced change in the site's charge state using a SET, rather than detecting the associated photocurrent.

Erbium's large electronic magnetic moment of its ground state, the $I$=7/2 nuclear spin of its 167 isotope along with the coincidence of the $^{4}$I$_{15/2}$ - $^{4}$I$_{13/2}$ transition with the 1.5 $\muup$m transmission window of silica optical fibers make Er centres appealing for quantum information applications\cite{kenyon_erbium_2005,vinh_photonic_2009,bertaina_rare-earth_2007}. The few existing studies in samples with a high concentration of Er have shown 0.1 s nuclear spin relaxation time\cite{baldit_identification_2010} and 100 $\muup$s electron spin dephasing time\cite{bertaina_rare-earth_2007}. Single-site detection grants access to low Er densities where one expects drastically enhanced coherence times in analogy to the impressive recent development for P in Si\cite{steger_quantum_2012}. The low emission rate from optically excited rare-earth ions such as Er$^{3+}$ makes pure optical detection of single sites challenging\cite{kenyon_erbium_2005,vinh_photonic_2009}. Recently though, the optical detection of a single rare-earth ion was demonstrated in a Pr-doped YAG nano-crystal\cite{kolesov_optical_2012}. The technique employed, involved the 2-step excitation of the ion to a high lying 5d-electron state and detecting the resultant emission\cite{kolesov_optical_2012}. It was conducted at room temperature and exhibited low detection efficiency and low frequency resolution making state readout not feasible.

The $^{4}$I$_{15/2}$ - $^{4}$I$_{13/2}$ transition is between states within the inner 4f-electron shell of the Er$^{3+}$ ion, which is well shielded from the surrounding lattice by filled outer shells, resulting in narrow spectral linewidths and the potential for high resolution frequency addressing. At liquid helium temperatures homogeneous linewidths as narrow as 50 Hz have been observed for the transition in Er$^{3+}$:Y$_2$SiO$_5$\cite{sun_recent_2002}. Prior to the present work there have not been any sub-inhomogeneous studies conducted on optical transition in Er centres in silicon. The observed lifetime of emission of 2 ms from the $^{4}$I$_{13/2}$ state for Er$^{3+}$ ions in silicon implies a minimum linewidth of 150 Hz\cite{priolo_excitation_1998}.

\begin{figure*}
\includegraphics[width=\textwidth]{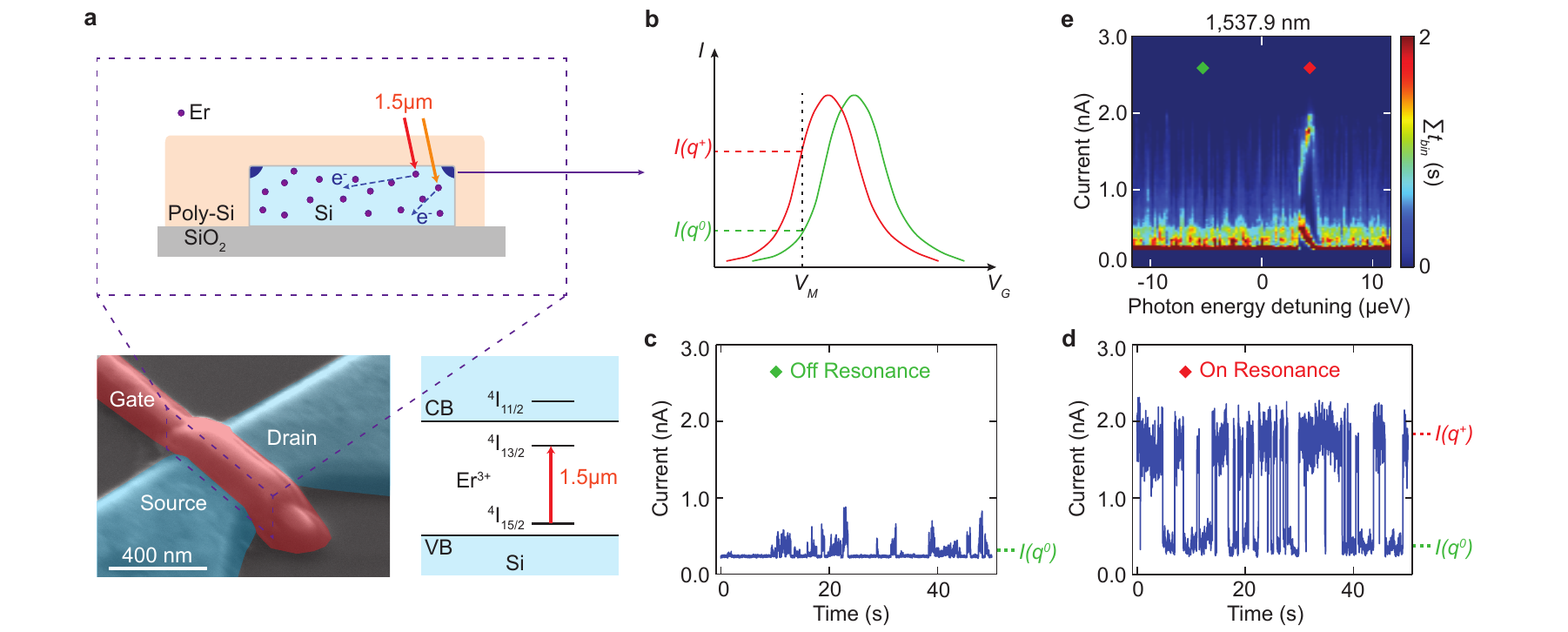}
\caption{\label{Fig:principles}Photoionization spectroscopy of an individual Er$^{3+}$ ion.
\textbf{a,} Coloured scanning electron micrograph of a typical SET device used in this study and a band structure of Er$^{3+}$ ions in silicon. Top blow-up: Schematic cross-section of the SET showing the optical addressing of individual Er$^{3+}$ ions.
\textbf{b,} The SET charge sensing scheme. The loss of an electron due to photoionization induces a transient shift of the $I$-$V_{G}$ curve towards lower gate voltage, causing a change in current from $I(q^{0})$ to $I(q^{+})$.
\textbf{c, d,} The current-time traces recorded with a fixed gate voltage ($V_{M}$) under (\textbf{c}) non-resonant and (\textbf{d}) resonant illumination.
\textbf{e,} The histogram of current-time traces as a function of the photon energy detuning. The photon energy of the illumination is detuned with respect to the centred wavelength of 1,537.9 nm.
}
\end{figure*}

The resonant photoionization of individual Er$^{3+}$ ions is studied in an Er-implanted SET (Fig. \ref{Fig:principles}a), which works as a charge sensor. The $^{4}$I$_{15/2}$ - $^{4}$I$_{13/2}$ transition of an Er$^{3+}$ ion has a relatively high probability, when a laser is tuned to its resonant wavelength, and the Er$^{3+}$ ion could be further ionized due to a second-photon process or an Auger process. The charge displacement induced by an ionization event simultaneously leads to a change in the tunnelling current of the SET. To get a high sensitivity, the SET is biased close to the degeneracy point between two charge states, i.e. at the edge of one Coulomb peak (Fig. \ref{Fig:principles}b). Accordingly, the transconductance is large, and a small charge displacement in the sensitive region will lead to a significant change in the tunnelling current\cite{pioda_single-shot_2011,hanson_spins_2007}. The photoionization of individual Er$^{3+}$ ions leads to a significant change in tunnelling current (Fig. \ref{Fig:principles}b). The $^{4}$I$_{15/2}$ - $^{4}$I$_{13/2}$ transition of each Er$^{3+}$ ion has a specific resonant photon energy, so individual Er$^{3+}$ ions are distinguished by the resonant photon energy. When the laser is tuned to a non-resonant wavelength, the tunnelling current mainly stays at the background level as shown in Fig. \ref{Fig:principles}c. In contrast, when the laser is tuned to a resonant wavelength of an Er$^{3+}$ ion, the photoionization of the Er$^{3+}$ ion leads to a rise in the tunnelling current, and then the current drops back due to its neutralization, contributing a two-level current-time trace (Fig. \ref{Fig:principles}d), which suggests only one single Er$^{3+}$ ion is ionized (details in Methods). Figure \ref{Fig:principles}e shows a photoionization spectrum of a single Er$^{3+}$ ion. Current-time traces are recorded at a series of photon energies, and then the histogram showing the distribution of current in time is plotted as a function of the photon energy detuning. The colour in Fig. \ref{Fig:principles}e represents the time ($\Sigma t_{bin}$) during which the current stays within one bin, and a 0.02-nA bin size is used for all the analysis.

As shown in Fig. \ref{Fig:principles}a, the SET has a Si channel wrapped with the gate. The SET is biased below the threshold voltage, so that the current tunnels through the corner regions of the Si channel\cite{sellier_subthreshold_2007}. Consequently, the charge sensor is more sensitive to the Er$^{3+}$ ions which are closer to the corner regions in the channel, and different Er$^{3+}$ ions have different capacitive coupling leading to different detection sensitivity. The change in current (Fig. \ref{Fig:principles}b-d) accords with the loss of an electron, indicating that it's due to the ionization of the Er centre, whereas the gain of an electron will lead to a shift opposite to that in Fig. \ref{Fig:principles}b. The small fluctuations in current, which we attribute to the trap states in the insulating layer or the oxide layer with weak capacitive coupling\cite{tettamanzi_interface_2011}, can be suppressed by a proper anneal before the device fabrication. The readout efficiency is mainly limited by the efficiency of the excitation from the $^{4}$I$_{13/2}$ excited state into the conduction band, which can be increased to close to 100\%, by increasing the intensity of the light used to drive this final ionization step. We observe resonances via the photoionization spectroscopy mostly between 1,535 nm and 1,539 nm, which is consistent with the $^{4}$I$_{15/2}$ - $^{4}$I$_{13/2}$ transition of Er$^{3+}$ ions in silicon\cite{kenyon_erbium_2005,vinh_photonic_2009}.

\begin{figure}[h]
\includegraphics[width=\columnwidth]{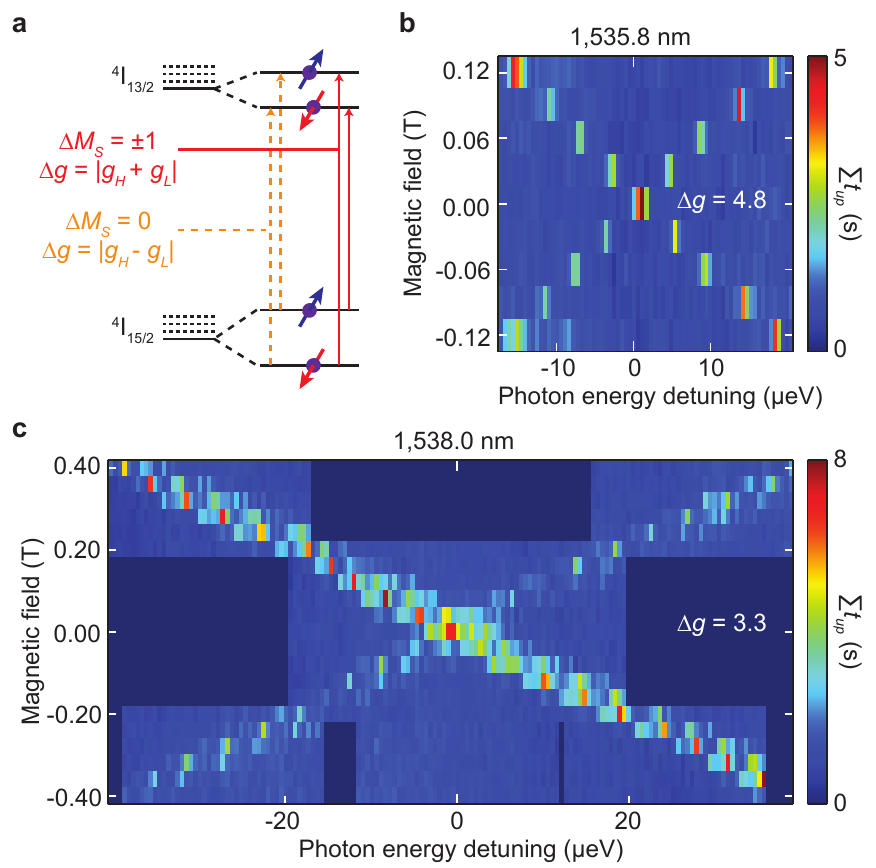}
\caption{\label{Fig:Zeeman}The Zeeman effect of individual Er$^{3+}$ ions.
\textbf{a,} Schematic diagrams showing the Zeeman splitting and optical transitions of Er$^{3+}$ ions in silicon. The splitting of the $^{4}$I$_{13/2}$ and $^{4}$I$_{15/2}$ states depends on the site symmetry of the Er centre.
\textbf{b,} The Zeeman splitting scan of the Er resonance with the centred wavelength of 1,535.8 nm. Each pixel stands for a current-time trace recorded for 50 s.
\textbf{c,} The Zeeman splitting scan of the Er resonance with the centred wavelength of 1,538.0 nm.
}
\end{figure}

In the next experiment, we study the Zeeman effect of individual Er$^{3+}$ ions, as the Zeeman effect is an essential tool to determine the site symmetry of Er centres. Er$^{3+}$ ions tend to take 3+ valence characteristic of the Si lattice, so the 4f electrons of Er$^{3+}$ ions have the ground state of $^4$I$_{15/2}$ and the first excited state of $^4$I$_{13/2}$\cite{kenyon_erbium_2005}. The degeneracy is lifted by the crystal field, so that each state splits into several levels depending on the symmetry of the Er centre\cite{kenyon_erbium_2005}. The transition between the lowest level of $^4$I$_{15/2}$ and the lowest level of $^4$I$_{13/2}$ is responsible for the strong emission band around 1.54 $\muup$m, and the Zeeman splitting of those two levels in the case of double degeneracy is shown in Fig. \ref{Fig:Zeeman}a. The doublet states can be described by an effective spin $S$=1/2, and the Zeeman interaction has the form: $H$=$\beta_e\mathbf{B}\cdot\mathbf{g}\cdot\mathbf{S}$, where $\beta_e$ is the electronic Bohr magneton, $\mathbf{B}$ is the magnetic field, and $\mathbf{g}$ is the $g$-factor matrix\cite{guillot-noel_hyperfine_2006}. The Zeeman splitting energy of the higher (lower) energy doublet is proportional to $g_H$ ($g_L$). As shown in Fig. \ref{Fig:Zeeman}a, the energy difference between the two $\varDelta M_S$=$\pm$1 ($\varDelta M_S$=0) transitions can be described by $\varDelta E = \beta_e\varDelta g B$, where $\varDelta g$ is the $g$-factor difference. In this study, we measured the Zeeman splitting of 4 spectrally isolated Er resonances, and observed the $g$-factor difference from 1.6 to 10.8.

Figure \ref{Fig:Zeeman}b,c shows the Zeeman splitting of Er$^{3+}$ ions. Current-time traces are taken at a series of photon energies and magnetic fields, and each pixel in Fig. \ref{Fig:Zeeman}b,c represents one current-time trace. When an Er$^{3+}$ ion is ionized, the current will exceed a certain threshold, which is determined by the background current fluctuation under non-resonant illumination. For each current-time trace, the time ($t_{up}$), during which the current exceeds the threshold, is integrated and gives the values ($\Sigma t_{up}$) plotted in Fig. \ref{Fig:Zeeman}b,c. As shown in Fig. \ref{Fig:Zeeman}b, the resonance shows up at the photon energy detuning of 1 $\muup$eV at zero magnetic field, and starts to split into two diagonal arms with increasing magnetic field. It is due to the Zeeman effect of one individual Er$^{3+}$ ion, with $\Delta g\approx 4.8$. Similarly, the Zeeman splitting of the resonance around 1,538.0 nm is studied as shown in Fig. \ref{Fig:Zeeman}c, and the rectangular regions denoted by the darkest blue colour are not scanned. There appear to be two resonances with similar resonant wavelengths and the same $g$-factor difference ($\Delta g\approx 3.3$) but with different signal intensity. This could be due to two individual Er$^{3+}$ ions with the same site symmetry but with different capacitive coupling. Furthermore, the Zeeman splitting of the resonance around 1,538.0 nm shows polarization dependence. As shown in Fig. \ref{Fig:Zeeman}c, the diagonal arm is weaker than the anti-diagonal one. By modifying the polarization of the light entering the cryostat, the diagonal arm was tuned to be stronger than the anti-diagonal one. The site symmetry of individual Er centres can be determined with the polarization dependence and a rotating magnetic field measurement. Spin selective excitation even for degenerate spin states can be achieved with the maximum contrast of the polarization dependence, which allows spin readout without a magnetic field.

\begin{figure*}
\includegraphics[width=\textwidth]{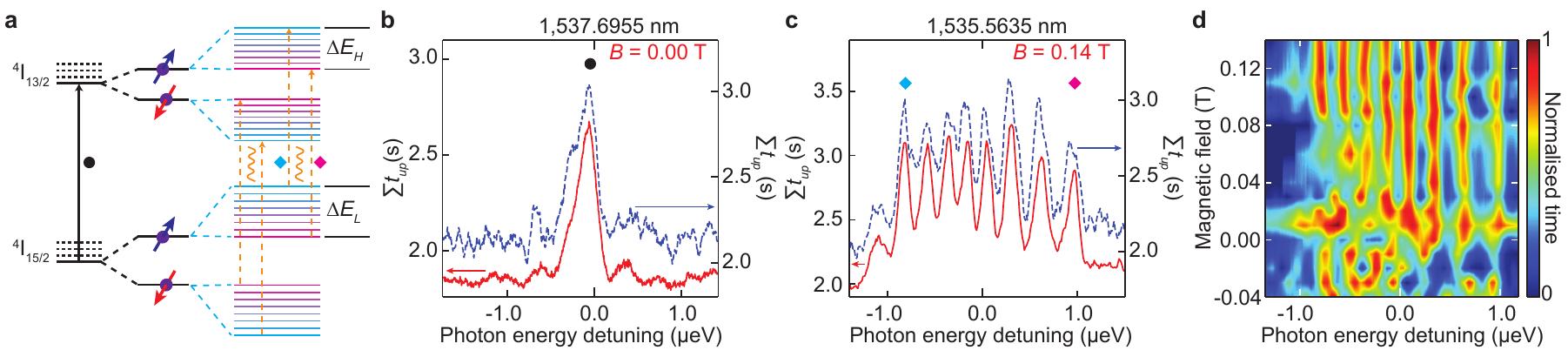}
\caption{\label{Fig:Hyperfine}The hyperfine structure of an individual Er$^{3+}$ ion.
\textbf{a,} Schematic diagrams showing the hyperfine splitting and the $\varDelta M_S$=0 transitions (orange dashed line) of $^{167}$Er$^{3+}$ ions at high magnetic field, and the optical transition of Er$^{3+}$ ions with zero nuclear spin at zero magnetic field for comparison (black solid line).
\textbf{b,} The photoionization spectrum of an Er$^{3+}$ control ion with zero nuclear spin. The red solid (blue dashed) line represents the data with (without) removing the broadening.
\textbf{c,} The photoionization spectrum of a single $^{167}$Er$^{3+}$ ion. The eight significant peaks correspond to optical transitions of the eight nuclear spin states of $^{167}$Er ($I$=7/2).
\textbf{d,} The contour plot of the photoionization spectra of the $^{167}$Er$^{3+}$ ion showing evolution of the hyperfine interaction.
}
\end{figure*}

The hyperfine structure is of great interest as the nuclear spin has long coherence times for quantum information storage\cite{steger_quantum_2012,hedges_efficient_2010,simmons_entanglement_2011}. In addition, it is a strong evidence for distinguishing between different ions as well as other defects. Erbium has six stable isotopes, among which only $^{167}$Er has a nonzero nuclear spin of $I$=7/2, leading to eight nuclear spin states. At high magnetic field, the hyperfine interaction can be treated as a perturbation of the Zeeman effect\cite{smith_hyperfine_1965}, so each electron spin state will split into eight sublevels due to the hyperfine interaction (Fig. \ref{Fig:Hyperfine}a). At low magnetic field, the hyperfine interaction is comparable to the Zeeman effect, so the sublevels will mix\cite{mcauslan_reducing_2012}.

In order to investigate the hyperfine structure of $^{167}$Er$^{3+}$ ions, we implanted $^{167}$Er and $^{168}$Er with zero nuclear spin as control ions. We first study the photoionization spectrum of an Er$^{3+}$ control ion with zero nuclear spin. The integrated time ($\Sigma t_{up}$) is plotted as a function of the photon energy detuning, as indicated by the blue dashed line in Fig. \ref{Fig:Hyperfine}b. The same spectral asymmetry as that in Fig. \ref{Fig:principles}e is observed. We attribute the asymmetry to the correlation between the Stark shift of the $^{4}$I$_{15/2}$ - $^{4}$I$_{13/2}$ resonance and a broadening of the Coulomb peak, both of which are sensitive to fluctuating electric fields in the channel. The fluctuating field is attributed to the laser excitation of trap states in or near the channel. Since we directly observe the effect on the Coulomb peak it is possible to remove part of this broadening of the peak. After applying this correction (details in Methods), a minimum FWHM spectral width of 50 neV is observed as indicated by the red solid line in Fig. \ref{Fig:Hyperfine}c. To significantly reduce the linewidth further it will be necessary to reduce the density of trap states. As well as the electric field induced shifts the line is expected to be broadened through magnetic interactions with $^{29}$Si up to tens of neV and with paramagnetic centres in the device. It is expected from analogy with observations in Er$^{3+}$:Y$_2$SiO$_5$\cite{sun_recent_2002} that applying a large magnetic field will suppress this broadening mechanism.

In the following, we show the hyperfine structure of one $^{167}$Er$^{3+}$ ion. As shown in Fig. \ref{Fig:Hyperfine}c, the photoionization spectrum taken at high magnetic field ($B=0.14$ T), reveals eight resonant peaks with the photon energy difference of about 0.2 $\muup$eV between each other. The high spectral resolution allows the nuclear spin readout with potential for single-shot readout and manipulating the nuclear spin states. As the addressability doesn't rely on a specific magnetic field, the photoionization spectra are measured at a series of magnetic fields, as shown in Fig. \ref{Fig:Hyperfine}d. The Zeeman shift is subtracted to show evolution of the hyperfine interaction. At high magnetic field, eight significant peaks are observed all through (0.08 T $\leqslant B \leqslant$ 0.14 T), as the hyperfine interaction can be treated as a perturbation of the Zeeman effect. At low magnetic field, multiple resonances show up (-0.04 T $\leqslant B \leqslant$ 0.06 T), revealing the mixing of the hyperfine sublevels, since the hyperfine interaction is comparable to the Zeeman effect.

The eight significant peaks, representing the eight different nuclear spin states of $^{167}$Er, demonstrate that the resonances are due to the $^{167}$Er$^{3+}$ ion rather than other ions or defects. These eight hyperfine peaks (Fig. \ref{Fig:Hyperfine}c) correspond to the allowed transitions ($\varDelta M_I$=0) preserving the nuclear spin states, but it's still a question whether they are due to the $\varDelta M_S$=0 or $\varDelta M_S$=$\pm$1 transitions. As shown in Fig. \ref{Fig:Hyperfine}a, we attribute them to the $\varDelta M_S$=0 transitions for two reasons. First, the energy difference between the two most distant hyperfine peaks is only about 1.7 $\muup$eV (Fig. \ref{Fig:Hyperfine}c), which is much smaller than the typical splitting energy of the $\varDelta M_S$=$\pm$1 transitions of Er$^{3+}$ ions. The electron paramagnetic resonance measurements of $^{167}$Er$^{3+}$ ions in crystals show a splitting energy ($2\varDelta E_L$) of about 30 $\muup$eV\cite{yang_electron_2009,bertaina_rare-earth_2007}, which corresponds to the $\varDelta M_S$=$\pm$1 transitions. Second, a ninth peak shows up beyond the region between the two most distant peaks (at the photon energy detuning of -1.1 $\muup$eV in Fig. \ref{Fig:Hyperfine}c). The ninth peak is much weaker than the eight peaks but still recognizable, which we attribute to a not-allowed transition. The energy of the not-allowed transitions ($\varDelta M_I$=$\pm$1) of Er$^{3+}$ ions can exceed the region between the two most distant peaks of the allowed transitions, only in the case of the $\varDelta M_S$=0 transitions. Consequently, the eight significant peaks are attributed to the $\varDelta M_S$=0 transitions, and the splitting energy is expressed as $|\varDelta E_H-\varDelta E_L|$=1.7 $\muup$eV.

Hybrid optical/electrical access to single spins of individual ions in a nano-transistor has been demonstrated, which is applicable for other defects in solids. Specifically, with an Er-implanted SET the photoionization spectroscopy allows real-time observation of single optical excitation events avoiding the bottleneck of photon collection. Furthermore, high-resolution optical frequency addressing circumvents the limitations due to thermal broadening in earlier electrical detection of impurity spins\cite{morello_single-shot_2010}. Our findings open the way to optically address and manipulate the electron and nuclear spin states of an individual defect in a solid beyond the nitrogen-vacancy centre in diamond. In addition, this hybrid optical/electrical technique boosts the microstructural study of ions in a semiconductor to a single-site level, including microscopic aspects, electrical and optical activity, etc.

An approach that combines dopant ions (e.g. Er, P) with quantum optical control and semiconductor fabrication technologies represents an attractive platform to realize a scalable quantum computation and communication architecture. Such a system could consist of individual ions inside a ring cavity coupled with each other via photons, and nearby charge-sensing devices used to read out the spin states of individual ions and to control the coupling between ions by Stark tuning. The ring cavities can be connected by optical waveguides, which enable quantum information transfer between individual ions in different ring cavities. 
Here we demonstrated the first step towards such a system, i.e. optical addressing of individual ions, and further improvement can be made by reducing the observed linewidth as discussed previously. However, there are essential questions to be addressed in the future, such as electron and nuclear spin coherence times of Er (P) ions, the influence of photoionization on nuclear spin coherence, and spin-photon entanglement.

\textbf{METHOD SUMMARY}

The devices are fabricated with the same technique as the previous study\cite{lansbergen_gate-induced_2008}. After complete device fabrication, an Er:O co-implantation (dose ratio 1:6 ) is performed with the implantation energy of 400 keV and 55 keV, respectively. There should be approximately 30-40 Er ions in the sensitive region of one Coulomb peak. Under the erbium implantation conditions we used, the beam is estimated to have been composed of 70-80\% $^{168}$Er and 20-30\% $^{167}$Er. The devices are then annealed at 700$^\circ$C in N$_{2}$ for 10 minutes to remove the implantation damage and to initiate the formation of Er centres. All the measurements are carried out in a liquid helium cryostat at 4.2 K. The laser beam, with 4-5 mW optical power, goes through a single-mode fiber and irradiates the sample with a diameter of about 1 mm. In the initial phase of the experiments (Figs. \ref{Fig:principles}\ and \ref{Fig:Zeeman}), a commercial tunable laser with an external cavity is used. To keep a high precision, we set one centred wavelength with the motor-actuator, and sweep the wavelength around the centred wavelength with the piezo-actuator. In the high-resolution experiments (Fig. \ref{Fig:Hyperfine}), the wavelength of another laser is stabilized to about 0.01 pm, and a wavelength meter is used to compensate the thermal drift.

\textbf{Acknowledgements} We thank R. Ahlefeldt, J. Bartholomew, R. Elliman, N. Manson and A. Morello for discussions. We also thank M. Hedges and T. Lucas for their help in the initial phase of the experiments. The devices were fabricated by N. Collaert and S. Biesemans (IMEC). The work was financially supported by the ARC Centre of Excellence for Quantum Computation and Communication Technology (CE110001027), and the Future Fellowships (FT100100589 and FT110100919).

\textbf{Author Contribution}
N.S. and J.C.M. designed and performed the implantation. C.M.Y., M.J.S. and S.R. designed and conducted the experiments. C.M.Y., M.R. and G.G.d.B. carried out the experiments. All the authors contributed to analysing the results and writing the paper. Correspondence and requests for materials should be addressed to S.R. (s.rogge@unsw.edu.au).

\bibliography{Er1-1new.bib}

\clearpage
\textbf{METHODS}

\textbf{Details of the devices.} The devices used in this study are n-p-n field-effect transistors with a polycrystalline silicon gate wrapped around the p-type silicon channel separated by the gate dielectric. The p-type channel has a boron doping of 10$^{18}$ cm$^{-3}$. After complete device fabrication, an Er:O co-implantation is performed with the implantation energy of 400 keV and 55 keV and the ion fluence of $4\times10^{12}$ cm$^{-2}$ and $3\times10^{13}$ cm$^{-2}$, respectively. This leads to an Er:O dose ratio of about 1:6 in the channel region. Under the erbium implantation conditions we used, the beam is estimated to have been composed of 70-80\% $^{168}$Er and 20-30\% $^{167}$Er. The presence of both oxygen impurities\cite{kenyon_erbium_2005} and boron impurities$^{31}$ is known to enhance the luminescence of the Er$^{3+}$ ions in silicon. The 700$^\circ$C post-implantation anneal is within the thermal processing window for Er centre activation in silicon$^{31}$.

In the experiments, the device is biased below the threshold voltage, and only the corner regions of the silicon channel go into inversion\cite{sellier_subthreshold_2007}. A peak of the $I$-$V_G$ curve is due to the Coulomb blockade in one of the two corner regions, where the current flows. The sensitive region is defined as the region, in which one elementary charge change can be detected, with taking the current noise and the transconductance of the Coulomb peak into account. The sensitive region of one Coulomb peak is estimated to be the corresponding channel-corner region with a dimension of $100\times 50\times 20$ nm (length $\times$ width $\times$ height) for the device shown in Fig. 1a. Simulations of the ion implantation based on SRIM$^{32}$ show that there should be approximately 30-40 Er ions in the sensitive region of one Coulomb peak.

\textbf{Experimental details and data analysis.} All the measurements are carried out in a liquid helium cryostat at 4.2 K. The laser beam, with 4-5 mW optical power, goes through a single-mode fiber and irradiates the sample with a diameter of about 1 mm. In the initial phase of the experiments (Figs. \ref{Fig:principles}\ and \ref{Fig:Zeeman}), a commercial tunable laser with an external cavity is used. To keep a high precision, we set one centred wavelength with the motor-actuator, and sweep the wavelength around the centred wavelength with the piezo-actuator. The current-time traces in Fig. \ref{Fig:principles}c and Fig. \ref{Fig:principles}d taken at two different photon energies are consistent with the photoionization spectrum as indicated by the green and red diamonds in Fig. \ref{Fig:principles}e, respectively. For instance, the current mainly stays at the background level (0.4 nA) at the photon energy detuning of -5 $\muup$eV, while the current jumps between two levels (1.8 nA and 0.4 nA) at the photon energy detuning of 4 $\muup$eV. It is worth to note that the two-level trace (Fig. \ref{Fig:principles}d) suggests only one single Er$^{3+}$ ion is ionized. Multiple ions with different capacitive coupling will lead to a current-time trace with more than two levels, while two ions with the same capacitive coupling will lead to a current-time trace with three levels once they are simultaneously ionized. We attribute the charge displacement to the ionization of an Er$^{3+}$ ion rather than the charge fluctuations of the trap states, based on the observation that all the Er$^{3+}$ ions that we observed contribute to a shift of the Coulomb peak towards lower gate voltage. In the high-resolution experiments (Fig. \ref{Fig:Hyperfine}), the wavelength of another laser is stabilized to about 0.01 pm, and a wavelength meter is used to compensate the thermal drift. The asymmetry as well as part of the broadening of the resonant peak is removed by adding an photon energy offset to the data, and then the time, during which the current exceeds the threshold, is integrated and gives the values plotted as the red solid line in Fig. \ref{Fig:Hyperfine}b,c. In comparison, the red solid line with removing the broadening shows smaller widths and less noise than the blue dashed line without removing the broadening, nevertheless, the resonances in the latter are still clearly visible, as shown in Fig. \ref{Fig:Hyperfine}b,c.

\textbf{References}

\noindent{$^{31}$ J. Michel, J. L. Benton, R. F. Ferrante, D. C. Jacobson, D. J. Eaglesham, E. A. Fitzgerald, Y.-H. Xie, J. M. Poate, and L. C. Kimerling, Journal of Applied Physics \textbf{70}, 2672 (1991)}

\noindent{$^{32}$ J. F. Ziegler, M. Ziegler, and J. Biersack, Nuclear Instruments and Methods in Physics Research Section B: Beam Interactions with Materials and Atoms \textbf{268}, 1818 (2010).}
\end{document}